\documentclass[preprint,sort,compress]{elsarticle}
\usepackage[]{geometry}
\usepackage{macros}
\usepackage{blindtext}
\usepackage[separate-uncertainty=true,multi-part-units=single,per-mode = symbol,list-units = single,product-units = repeat,range-units = single,tight-spacing = true]{siunitx}
\usepackage{graphicx}
\usepackage[export]{adjustbox}
\usepackage[colorlinks]{hyperref}
\usepackage{amsmath}
\usepackage{amsfonts}
\usepackage{accents}
\usepackage{amsxtra}
\usepackage{amstext}
\usepackage{amssymb}
\usepackage{latexsym}
\usepackage{dsfont}
\usepackage{textcomp}
\usepackage[normalem]{ulem}
\usepackage{multirow}
\usepackage{color}
\usepackage{epstopdf}
\usepackage{caption, subcaption}
\usepackage{bm}
\usepackage{soul}
\usepackage{natbib}
\usepackage{hyperref}
\usepackage{upgreek}
\usepackage{lscape}
\hypersetup{
	colorlinks=true,
	linkcolor=cyan,
	filecolor=magenta,      
	urlcolor=blue,
}
\hypersetup{pdfauthor={Name}}

\begin{document}

\begin{frontmatter}
	
\title{Time domain analysis of locally resonant elastic metamaterials under impact}

\author[1]{Erdem Caliskan}
\author[2]{Willoughby Cheney}
\author[2]{Weidi Wang}
\author[3]{Thomas Plaisted}
\author[2]{Alireza V. Amirkhizi}
\author[1]{Reza Abedi\corref{cor1}}
\ead{rabedi@utk.edu}

\affiliation[1]{organization={Department of Mechanical, Aerospace and Biomedical Engineering},
    addressline={University of Tennessee Knoxville, 1506 Middle Drive},
    city={Knoxville},
postcode={37916},
    state={TN},
    country={USA}}
\affiliation[2]{organization={Department of Mechanical Engineering},
    addressline={University of Massachusetts Lowell, 1 University Ave},
    city={Lowell},
    postcode={01854},
    state={MA},
    country={USA}}
\affiliation[3]{organization={DEVCOM Army Research Laboratory},
    addressline={Aberdeen Proving Ground},
    city={Aberdeen},
    postcode={21005},
    state={MD},
    country={USA}}

\cortext[cor1]{Correspondingauthor}

\begin{abstract}

The microstructure of a material can be engineered to achieve unique properties not found in nature. Microstructured materials, also known as metamaterials (MMs), can exhibit properties utilizing local resonance and dynamics of their heterogeneous microstructure that are activated below the traditional Bragg limit. In this study, the linear dynamic response of a low-frequency resonant ceramic MM slab is analyzed using the Finite Element Method (FEM) in the time domain. The MM is compared to monolithic slabs and other microstructured designs in terms of stress wave mitigation, peak load retardation, and energy transfer. Simulations are conducted using various boundary conditions and domain sizes to evaluate their influence on the performance. Potential graded slab designs and material damping effects are also discussed and are both shown to reduce the energy transmitted from the impact surface to the opposing surface significantly. The results showed that the MM slabs had superior performance in reducing the peak stress wave and reducing the transfer of energy. This study demonstrates that resonant ceramic MMs are a promising material design with unique and tunable properties that can be used for stress wave mitigation and structural protection applications. 

\end{abstract}

\begin{keyword}
Mechanical metamaterial, wave propagation, energy mitigation, impact, blast
\end{keyword}
\end{frontmatter}

\section{Introduction}\label{sec:introduction}

The dynamic properties of the materials, among other things, depend on their microstructure. Metamaterials (MMs) are described as materials with engineered microstructures that possess properties not found in nature. In particular, dynamic mechanical metamaterials may interact with stress waves in unprecedented ways. These structures are designed to seek control of the trajectory and dissipation characteristics of the applied waves \cite{srivastava_evanescent_2017}. Their application areas include but are not limited to blast protection \cite{tan_blast-wave_2014}, acoustic wave mitigation \cite{ding_two-dimensional_2010}, seismic shielding \cite{brule_experiments_2014}, subwavelength imaging \cite{moleron_acoustic_2015}, and lenses \cite{lv_metamaterial_2019}. Elastic/acoustic MMs can possess exotic overall mechanical properties. These properties include effective dispersive (frequency-dependent) and possibly negative effective mass density \cite{sheng_dynamic_2007, huang_wave_2009} or elastic moduli \cite{lv_metamaterial_2019} tensors. Mechanical MMs have attracted substantial attention from the scientific community due to their potential to achieve higher performance than traditional designs in elastic wave mitigation with these exotic properties. 

Compared to conventional design approaches, the engagement of microstructure in MMs can significantly enhance wave dissipation, thereby enabling more effective attenuation of stress wave energy. The frequency range in which an infinitely periodic microstructure does not allow wave propagation is called a bandgap \cite{brillouin_wave_1946}. Two mechanisms can create bandgaps: Bragg scattering and local resonance \cite{liu_wave_2012}. Bragg scattering bandgap can be achieved by symmetry, periodicity, and order in the structure \cite{hussein_dispersive_2007, hussein_dynamics_2014}. However, the bandgap frequencies for these structures are limited to approximately $c/a$ where $c$ is the wave speed, and $a$ is the lattice constant \cite{liu_three-component_2002}. This limitation renders the Bragg scattering approach not very attractive for mechanical MMs due to long wavelengths requiring very large structures \cite{hirsekorn_small-size_2004, wang_one-dimensional_2004}. On the other hand, locally resonant MMs can lower the bandgap frequency by several orders of magnitude since it is driven by the resonance frequency of the unit cells \cite{liu_locally_2000}.

Furthermore, MM arrays with local resonant structures can be further tailored using graded designs \cite{cummer_one_2007} as they do not require perfect periodicity. In other words, the wave speed profile of a finite array can be modified by changing the material properties or the geometry at the cell level \cite{amirkhizi_continuous_2017}. Thus, a slab (1D array with periodicity in the other two directions) can be modified to be effective at a wider frequency range and dissipate more energy. Beyond 1D arrays, researchers demonstrated that, theoretically, cloaking could be achieved with continuous grading effective density \cite{cummer_one_2007}. Moreover, locally resonant materials can exhibit ``damping emergence'' or ``metadamping'', which can increase the reduction, mitigation, or absorption of shocks \cite{hussein_metadamping_2013}. It must be noted that all physical realization should include loss and damping. In fact, it can be shown that the inclusion of damping in the analysis of even linear systems could provide physical insight to the point of explicit calculation of discontinuous features in the dispersive behavior of idealized lossless systems from continuous lossy behavior through a limiting process \cite{abedi_use_2020}. 

An early example of the application of locally resonant MMs was introduced as ``metaconcrete'', which showed substantial potential in wideband blast-wave mitigation owing to its complex dispersive response \cite{mitchell_metaconcrete_2014, mitchell_investigation_2015}. Furthermore, the fracture response of the metaconcrete was investigated through experimental and numerical studies \cite{mitchell_effect_2016, briccola_experimental_2016, kettenbeil_experimental_2018}. However, metaconcrete has limitations due to its design. First, it is designed to tackle blast, not ballistic or impact loadings. The wave mitigation properties of the MMs, such as energy absorption, are effective at specific bandgap frequencies. In addition, due to the nature of the designed cell, the coating material should be compliant, which diminishes its mechanical integrity. Finally, the overall weight is heavy for most applications. Another design recently revealed to mitigate impact energy is called hybrid MM due to a combination of local resonant and negative Poisson's ratio features \cite{zhou_impact_2022}. In this work, researchers focus on \emph{Frequency Domain} (FD) analysis and experimental investigation of the slab. However, they did not consider projectile impact and \emph{Time Domain} (TD) analysis. Thus, although MMs have been shown to potentially improve the dynamic response of structures, some of the issues mentioned above, such as high mass and low mechanical integrity, need to be addressed. Furthermore, for a more realistic evaluation of their performance, TD analysis of finite structures (in contrast with infinitely periodic ones) should be performed. 

This paper studies an H-shaped cellular structure, which is a variation of the promising cell design developed in \cite{Aghighi2019}. The design is targeted to be produced using emerging additive manufacturing techniques while using materials,\eg ceramics, which will provide significantly high strengths. This unit cell design may address the issues experienced in other resonant MMs \cite{mitchell_metaconcrete_2014, mitchell_investigation_2015} thanks to its seamless transition from enclosure to the resonator. Similarly, a monolithic-microstructured hybrid component can be designed owing to this manufacturing method, which can help protect the microstructure from direct interaction with a projectile. Furthermore, bandgaps can be tuned for ballistic and impact events since they can be optimized for very low frequencies of less than 5 kHz.  This is attained by the low natural resonance frequency of the oscillatory resonator.  
The recent progress in computing and design techniques can be advantageous for this type of cell topology. \cite{Wang2023} investigated the band structure of this type of unit cell and created a reduced-order model (ROM) that can successfully compute both the eigenfrequency and the time domain response with very high accuracy and computational speed. Furthermore, \cite{morris_optimizing_2023} used a genetic algorithm with ROM to optimize bandgaps of the H-cell by grading the slab through its thickness.

Here, we systematically investigate the H-cell MMs and their interaction with linear elastic waves, through TD analysis of finite arrays, aim to establish a fundamental understanding that may lead to further improvements in this area. Thus, linear dynamic loading with various boundary conditions is investigated. Stress wave mitigation using microstructured media is explored and compared with alternative unit cells considering peak load retardation and energy transfer. MM arrays are simulated under different loading types and structural configurations in the form of graded designs. Further investigation into the energy transfer relative to a same-size monolithic slab has been performed with the added effect of structural damping.

The organization of this paper is as follows. First, the details of the FEM simulations, including geometry and material properties, boundary conditions, as well as comparison geometries and metrics, are laid out in Section \ref{sec:methodology}. The results and discussion are divided into seven subsections. To begin with, the response of the slab is investigated using snapshots from the FEM TD simulation and energy analyzed in FD. Then, four MM slabs with different numbers of unit cells are compared in Section \ref{sec:results_domain_size}, to evaluate the array size effect. Here, the best-performing slab size is selected to be used in the following sections. Next, MM and monolithic slabs are compared, considering traction-free and transmitting boundary conditions to represent idealized back face configurations in Section \ref{sec:results_bc}. Herein, the differences in the response are laid out, and the transmitting boundary condition is selected for the remaining sections. After determining the domain size and the boundary conditions, the MM slab is compared with comparison non-resonating slabs under impact in \ref{sec:results_comparison}. Then, two heuristically designed graded slabs are compared with the uniform slab in Section \ref{sec:results_heuristic}. Finally, the effect of damping is discussed in \ref{sec:results_damping}. The paper concludes with the summary and conclusions along with the future outlook in Section \ref{sec:summary}, where the effectiveness of the MMs in stress wave mitigation is highlighted.

\begin{figure}[t] 
\centering
    \begin{subfigure}[b]{0.2\textwidth}  
    \centering 
    \includegraphics[width=\textwidth]{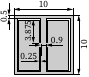}
    \caption[]%
    {}\label{fig:cell-MM}
    \end{subfigure}
    \begin{subfigure}[b]{0.2\textwidth}  
    \centering 
    \includegraphics[width=\textwidth]{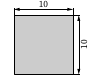}
    \caption[]%
    {}\label{fig:cell-monolithic}
    \end{subfigure}
    \begin{subfigure}[b]{0.2\textwidth}  
    \centering 
    \includegraphics[width=\textwidth]{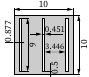}
    \caption[]%
    {}\label{fig:cell-equal}
    \end{subfigure}
    \begin{subfigure}[b]{0.2\textwidth} 
    \centering 
    \includegraphics[width=\textwidth]{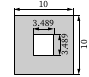}
    \caption[]%
    {}\label{fig:cell-square}
    \end{subfigure}
    \caption{Unit cells used in this study: (\subref{fig:cell-MM}) resonant MM cell, (\subref{fig:cell-monolithic}) monolithic cell, (\subref{fig:cell-equal}) same mass with equal stress path gap, and (\subref{fig:cell-square}) same mass with a square cavity.}\label{fig:comparison-cells}
\end{figure}

\begin{figure*}[t]
    \centering \includegraphics[width=0.8\textwidth]{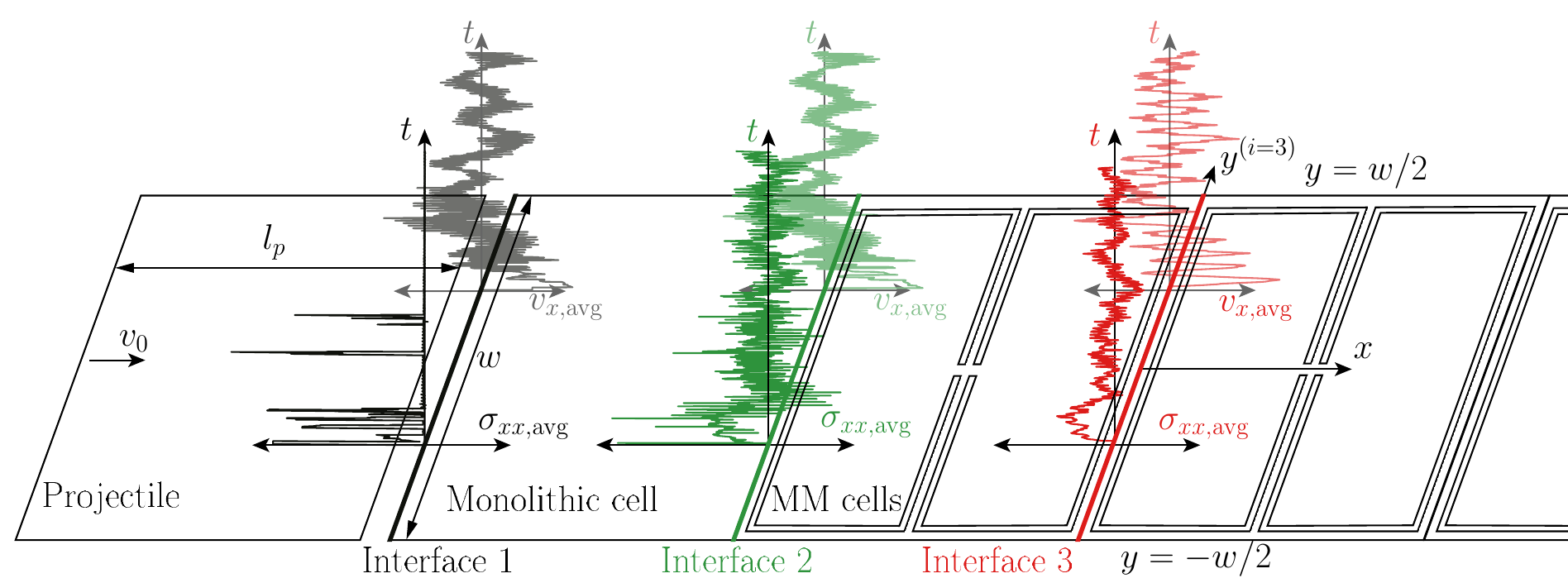}
    \caption{Schematic representation of interfaces at the slab with average stress and velocity values plotted against time. Metamaterial slabs have protective monolithic cells on both ends.}\label{fig:interfaces}
\end{figure*}

\section{Material and methods}\label{sec:methodology}

The material for the MM is alumina with a density of \qty{3985}{\kg\per\m\cubed}, the elastic modulus of \qty{300}{\GPa}, and Poisson’s ratio of \num{0.27}.
 Additive manufacturing can be used to print a target microstructural design.
The bandgap of the unit cell has been tuned between \qtyrange{20}{40}{\kHz} frequency range.  

The geometry of the square MM design is given in \Fig \ref{fig:cell-MM}. The resonator is H-shaped and connected to the frame from the middle horizontal line on the top and bottom sides. The area of the unit cell is \qtyproduct{10x10}{\mm}, while the wall, web, gap, and H-head dimensions are \qtylist{0.5;0.9;0.25;3.875}{\mm}, respectively. Three other candidate geometries are selected for comparison: monolithic, equal gap, and square cavity, shown in \Figs \ref{fig:cell-monolithic}, \ref{fig:cell-equal}, and \ref{fig:cell-square}, respectively. All comparison cells have the same outer dimensions, while the equal gap and square cavity also have the same mass as the H-cell MM. The monolithic cell represents the conventional design for protection. On the other hand, the equal gap is designed to have the same section area for stress waves to travel, while the square cavity is a simple cell design with the same mass. Long wavelength wave speeds ($c$, based on overall density and quasi-static modulus) for these unit cell designs are provided in Table \ref{tab:wavespeed}.

\begin{table}[h]
\centering
\caption{Long wavelength wave speed ($c$) for the unit cells. }\label{tab:wavespeed}
\begin{tabular}{l|l}
Unit cell  & $c$ (\qty{}{\m\per\s}) \\ \hline
MM         & 2951                      \\
Monolithic & 8677                      \\
Equal gap  & 3788                      \\
Square gap & 8603                     
\end{tabular}
\end{table}

Computational experiments are conducted using LS-DYNA \cite{hallquistLSDYNATheoryManual2006}. Postprocessing of the results is conducted by a Python script using the LASSO and qd libraries \cite{lasso_2023,diez_qd_2018}. The structure is meshed with uniform quadrilateral elements with a size of \qty{0.025}{\mm}. For the projectile impact simulations, the material is selected as cold-rolled steel, with a density, Young's modulus, and Poisson's ratio of \qty{7980}{\kg\per\m\cubed}, \qty{200}{\GPa}, and \num{0.3}, respectively. Contact between the projectile and ceramic slab is modeled using the mortar contact algorithm. Possible contacts inside the unit cell are also tracked, although no such contact has been detected for the simulations done in this study.

Quantitative comparisons are conducted by spatial averaging of the quantities of interest over the interface areas between unit cells (and same locations for the monolithic slab), and tracking the temporal evolutions of these average quantities. Interfaces numbering, along with example average stress and velocity quantities, are illustrated in \Fig \ref{fig:interfaces}. Here, the average stress and velocity values are plotted in two different time axes for the first three interfaces. $l_p$ and $v_0$ are the projectile length and initial speed, respectively, while $w$ is the width, which is fixed to \qty{10}{\mm}. 
Traction at a node as a function of $\mathrm{y}$-coordinate and the time, $t$ is calculated by,
\begin{equation}
    \vc{T^{(n)}}(y,t) = \tnssigma (y,t)  \cdot \vc{n},
\end{equation}
where $\tnssigma$ is stress interpolated at the nodes and $\vc{n}$ is the normal vector. 
Then, the energy flux density per unit time and spatial area, \ie power, ($\mathcal{P}$) and the energy transfer ($\mathcal{E}$) at the interface are calculated as:

\begin{align}
    \mathcal{P}(y,t) &= \vc{T^{(n)}}(y,t) \cdot \vc{v}(y,t),\label{eq:P}\\
    \mathcal{E}(t) &= \int_{0}^{t}\int_{-w/2}^{w/2} \mathcal{P}(y,t) \,\ensuremath{\mathrm{d}} y \,\ensuremath{\mathrm{d}} t.
\end{align}

The results are plotted as stress and energy at the interfaces between unit cells. Two metrics have been used to evaluate the performance: energy mitigation and slowdown of waves. The \emph{energy transfer ratio} is obtained using the ratio of the energy transfer of the last interface to that of the first one. The \emph{half-time} quantifies the time that it takes for the energy transfer to reach half of its peak value at each measurement location and is used as a measure of wave slow-down. These values have been used to compare the MM slab with the monolithic and the other comparison slabs. 

TD results are turned into FD by applying the Fourier transform (FT) to the spatial averages of stress and velocity. The FT of energy flux density $\mathcal{P}$ is obtained from the Poynting relation as:
\begin{equation}\label{eq:Poyning}
\hat{\mathcal{P}}(y,\omega) = 0.5 \mathrm{Re}\left[ 
\hat{\vc{T^{(n)}}}(y,\omega) \cdot \hat{\vc{v}}^*(y,\omega)\right],
\end{equation}
where $\hat{f}$ and $z^*$ stand for the Fourier transform of function $f$ and complex conjugate of number $z$, respectively, and $\mathrm{Re}$ is the real operator. The angular frequency is denoted by $\omega$. We apply the Poynting relation to the average traction and the conjugate of the average of velocity vectors to analyze the frequency content of the energy transfer.

\begin{figure}[t]
    \centering
    \includegraphics[width=0.45\textwidth]{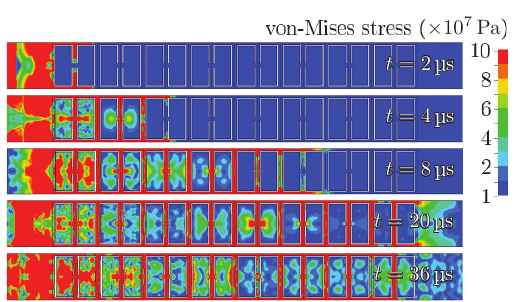}
    \caption{Snapshots of the von-Mises stress plotted on the array at five different timesteps.}\label{fig:cell_response_time_series}
\end{figure}

\begin{figure}[t]
    \centering
    \includegraphics[width=0.45\textwidth]{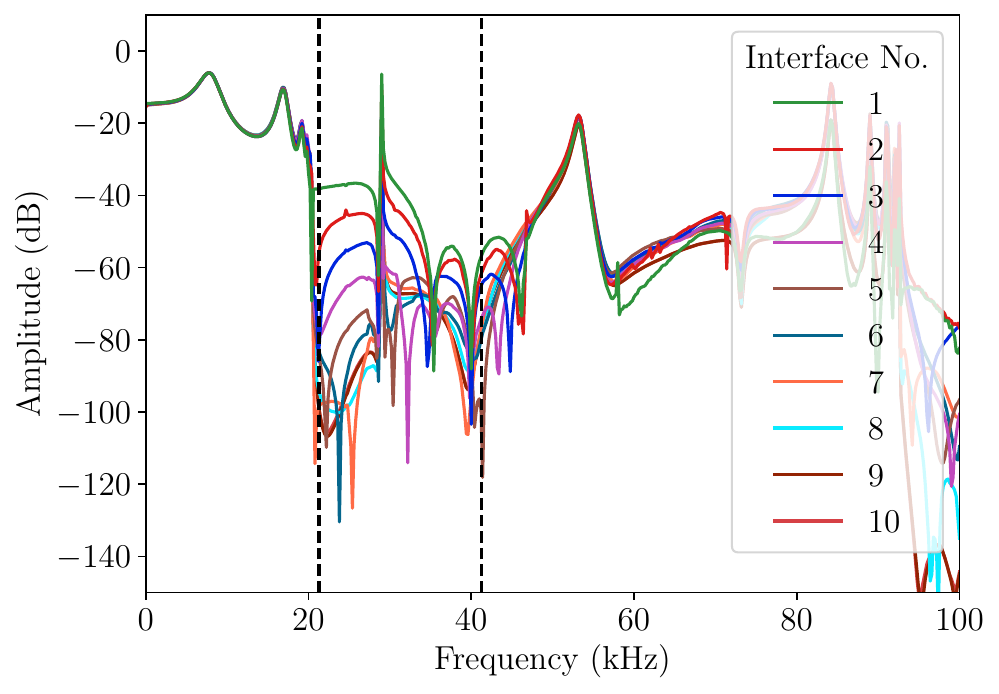}
    \caption{FT of the energy transfer. Dashed lines represent the boundaries of the FD calculated bandgap of the MM.}\label{fig:fft_energy}
\end{figure}

\begin{figure*}[ht!]
    \centering
    \includegraphics[width=0.99\textwidth]{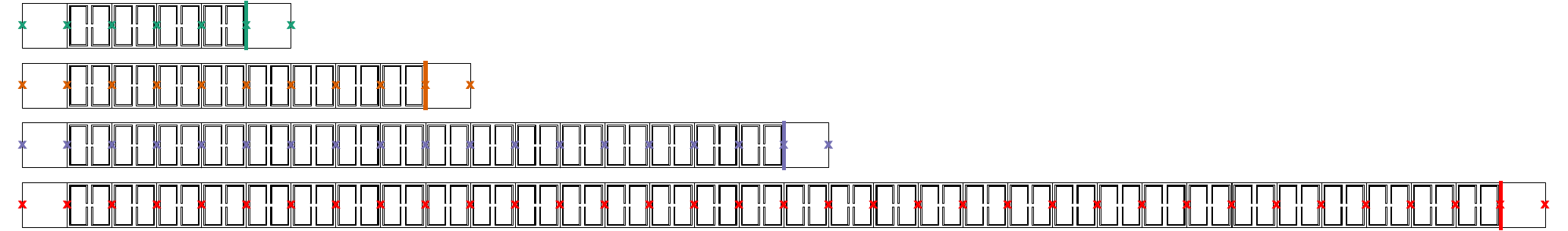}
    \caption{Four different-sized slabs were used in this study. Interfaces are marked with $\mathrm{x}$, and the last interface is marked with a solid line.}\label{fig:ds_slabs}
\end{figure*}

\begin{figure}[ht!]
    \centering
    \begin{subfigure}{0.35\textwidth}
        \includegraphics[width=\textwidth]{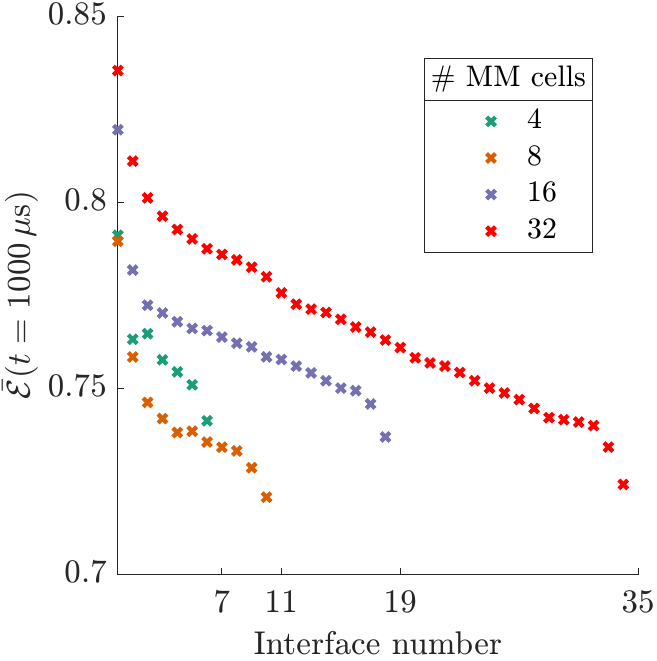}
        \caption{}\label{fig:ds_final}
    \end{subfigure}
    \begin{subfigure}{0.35\textwidth}  
        \includegraphics[width=\textwidth]{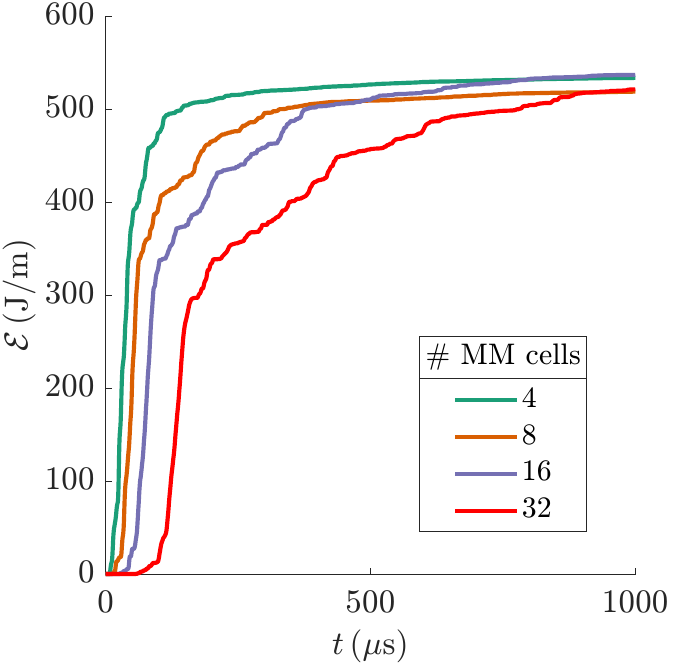}
        \caption{}\label{fig:ds_terminal}
    \end{subfigure}
    \caption{Domain size comparison. (\subref{fig:ds_final}) Energy transfer plotted against time at the final resonant cell interface for \numlist{4;8;16;32} cell slabs. (\subref{fig:ds_terminal}) Energy transfer magnitude at \qty{1000}{\us} in between cell interfaces for \numlist{4;8;16;32} cell slabs, normalized by initial projectile energy.}\label{fig:domain_size_results}
\end{figure}

Idealized boundary conditions, such as prescribed traction (PT) and prescribed velocity (PV), can be used to represent the two 
extremes of the impedances at the loaded front face; PT would provide more energy to the slab with a lower effective impedance, as the traction is provided from the PT boundary condition, whereas the corresponding normal velocity increases inversely proportional to the impedance. Since an MM slab has a lower impedance than a monolithic slab, it may observe several times higher energy transfer from the PT boundary condition. Yet, under PV, the opposite holds true. Nevertheless, modeling the projectile impact (PI) condition best represents actual impact scenarios, and blast loading (discussed below) is also a typical loading condition for these applications \cite{mitchell_metaconcrete_2014}. 

For the PI, the projectile length influences the time scale of the loading. While the default projectile length is $l_p 
= $\qty{20}{\mm}, when the effect of loading time 
 scale is examined, the projectile length of 
$l_p = \qtylist{2.5}{\mm}$ is also considered.
A square pulse profile is used for PT and PV loadings. 
 The duration of the square pulse $t_d$ takes the value of \qtylist{25}{\us}. Finally, 
the modified Friedlander approximation \cite{mitchell_metaconcrete_2014} is used to represent blast loading,
\begin{equation}
p(t)=p_0+P(1-\tau) e^{(-b \tau)}, \quad \tau=\frac{t-t_r}{t_d},\quad t>t_r,
\end{equation}
where $p(t)$ is the overpressure applied as a normal traction loading, $p_0$ is the surrounding hydrostatic air pressure, $p_0 + P$ is the initial overpressure, $r$ is the distance to the explosion, $t=t_r$ is the time for pressure front to reach the position $r$,
$t_d$ is the duration of the positive overpressure phase, $b$ is a decay parameter, and $\tau$ is a non-dimensional time scale. The time scale $t_d$ takes the same value used for square pulse PT and PV boundary conditions.

Since the material response is assumed to be linear in this study, the loading amplitude has a secondary importance. We use  $v_0 = \qty{30}{\m\per\s}$ for the PI loading. This results in an initial normal traction of \qty{600}{\MPa}. The load amplitudes for the PT, PV, and blast loadings are chosen to result in the same initial normal traction value.

\section{Results and Discussion}\label{sec:results}

In addition to the resonant dynamics of each cell, wave propagation characteristics in the microstructured media are affected by the domain size, boundary conditions, projectile size (for PI loading), and material properties. We will systematically investigate these parameters to compare the MM and other slab designs.

\subsection{TD and FD response of the MM}\label{sec:results_MM_response} 

The von-Mises stress fields plotted on the slab at different time values are illustrated in \Fig \ref{fig:cell_response_time_series}. High-resolution visualization of the stress is accomplished by using very small elements. First, after the projectile impact, most of the stress wave reflects from the first MM cell's free surface while the rest travels through the outer walls. Then, the resonators of the MM array start to engage with the wave. Cell engagement speed is much lower than the wave speed of the material. This behavior reduces the effective energy transfer speed of the slab. The reduction in downstream cell engagement can be seen in later steps by comparing the overall von-Mises stress magnitude inside the resonators.

The FD response analysis, represented by plotting the Fourier transform (FT) of energy flux density,\eqref{eq:Poyning} applied to average stress and conjugate of average velocity vectors, at each interface in \Fig \ref{fig:fft_energy}, is evaluated next. Here, we focused on \qtyrange{0}{100}{kHz} range. The FD bandgap for this design is from \qtyrange{20}{40}{kHz}. This plot shows that energy magnitude in all interfaces is nearly identical below \qty{20}{kHz}. Above this value, a consistent decrease in the magnitude of energy transfer at each progressive interface is observed. The magnitude levels get close to each other again after the upper bound. In addition, a peak is observed at \qty{30}{kHz}, which is called truncation resonance and recently explained in \cite{rosa_material_2023}. In general, the FT analysis of the TD results clearly matches the expected FD analysis for such a structure. 

\subsection{Domain size}\label{sec:results_domain_size} 

First, the effect of the domain (array) size on the wave propagation is investigated to select one size for further analysis. Since the frequency signature of the wave changes after each unit cell, increasing the number of cells will affect the total energy attenuation in a manner not proportional to the array length. Four slabs are analyzed with \numlist{4;8;16;32} resonant cells, given in \Fig \ref{fig:ds_slabs}, and loaded with the projectile of length $l_p = \qtylist{20}{\mm}$ while the back face is kept traction-free. In this figure, the final resonant interface is colored to represent the interfaces where the comparison is conducted.

Energy transfer at the final interface for each slab is represented in \Fig \ref{fig:ds_final}. The energy transfer requires different times to reach a plateau. This time correlates with the domain size. Regarding the absolute final magnitude of energy transfer, \num{8} cell slab is more efficient at reducing the transmitted energy than the others. Since this domain size is small, the energy transfer is faster to reach the plateau. The difference between the terminal values of the \num{4} and \num{8} cells is around \qty{3}{\percent}. The reflections inside the domain cause the \num{4} cell slab to have a slightly higher terminal value than the \num{8} cell slab. Energy transfer in the \num{16} cell slab is higher than \num{8} cell slab after a time at the final interface, reflecting as the crossing in the \Fig \ref{fig:ds_final}, is due to the interaction between length scale of the projectile and the total size of the slab.

To compare the performance throughout the domain, the energy transfers of all interfaces of each slab length are compared in \Fig \ref{fig:ds_terminal} at time 
\qty{1000}{\us}. The energies are normalized by the initial projectile kinetic energy, which is represented by $\bar{\mathcal{E}}$. The projectile maintains a nonzero velocity after the impact, explaining the normalized energy transfer values of less than one, ranging from \numrange{0.78}{0.83}, for the first interfaces. The design that reduces energy transmission the most is \num{8} cells with \qty{72}{\percent} energy attenuation at the final interface relative to the initial kinetic energy of the projectile. Since the \num{8} cell slab performs the better from the two perspectives above, we restrict the subsequent studies to  \num{8} cell slabs. 

\subsection{Back face boundary conditions}\label{sec:results_bc} 

\begin{figure}[t]
    \centering
    \begin{subfigure}{0.35\textwidth}
        \includegraphics[width=\textwidth]{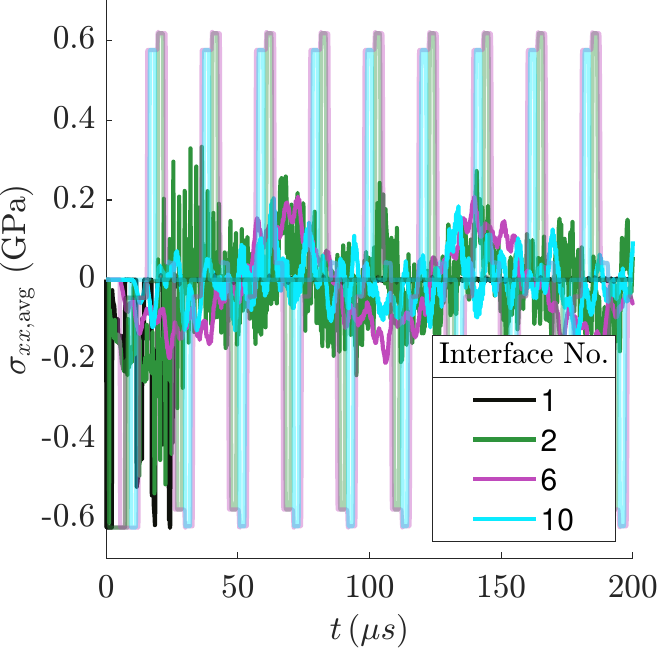}
        \caption{}\label{fig:projectile-traction-stress}
    \end{subfigure}
    \begin{subfigure}{0.35\textwidth}  
        \includegraphics[width=\textwidth]{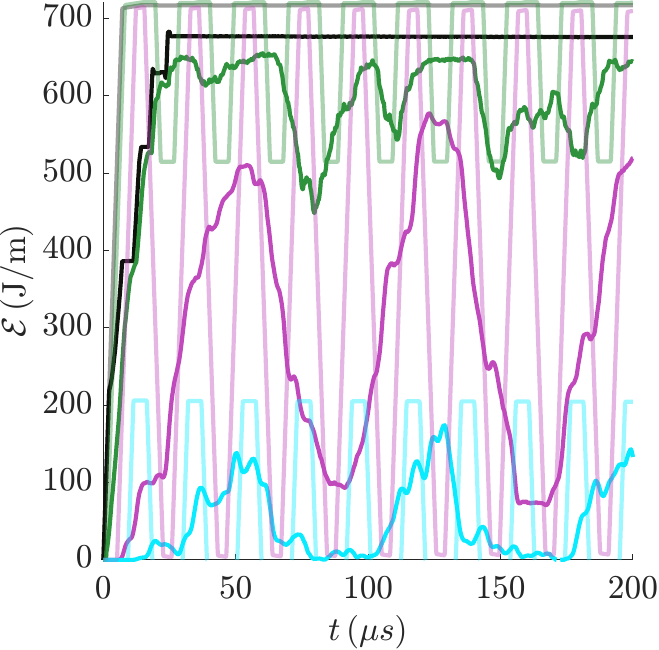}
        \caption{}\label{fig:projectile-traction-energy}
    \end{subfigure}
    \caption{Comparison of (\subref{fig:projectile-traction-stress}) average stress and (\subref{fig:projectile-traction-energy}) energy transfer as functions of time at different interfaces for MM (solid) and monolithic (transparent) slabs with traction-free boundary condition on the back-face under projectile impact.}\label{fig:projectile-traction-MM-monolithic}
\end{figure}

\begin{figure}[t]
    \centering
    \begin{subfigure}{0.35\textwidth}
        \includegraphics[width=\textwidth]{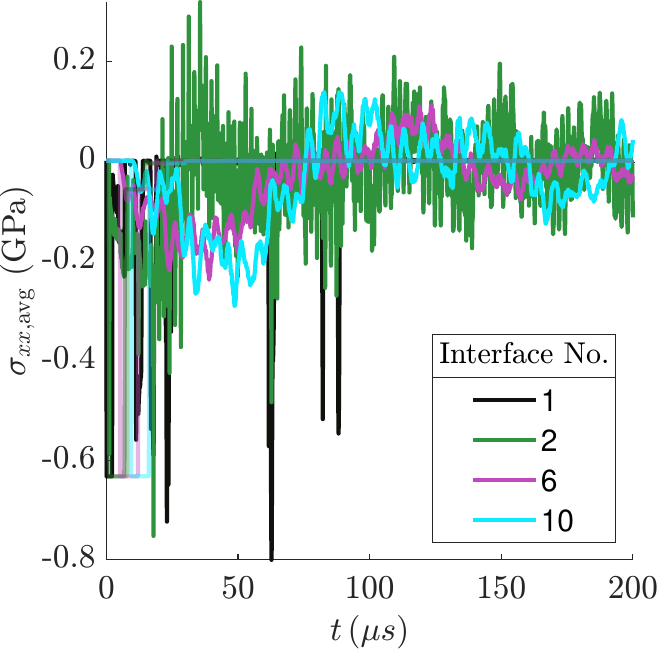}
        \caption{}\label{fig:projectile-transmitting-stress}
    \end{subfigure}
    \begin{subfigure}{0.35\textwidth}  
        \includegraphics[width=\textwidth]{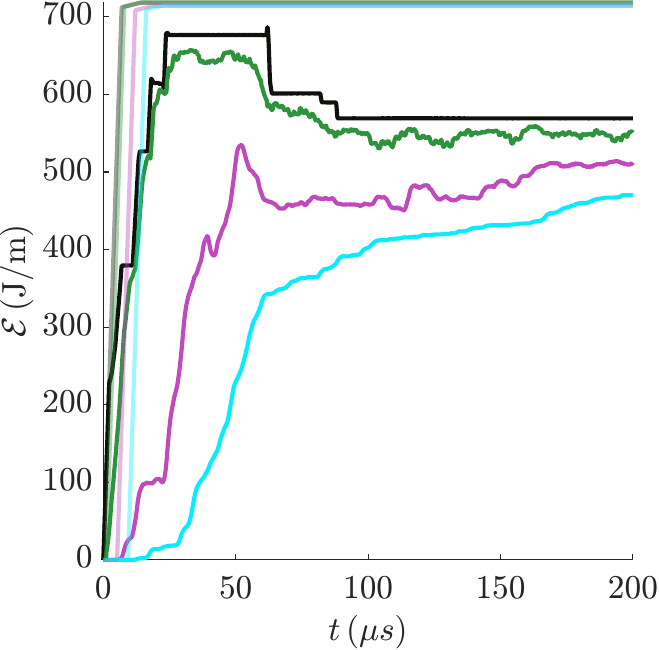}
        \caption{}\label{fig:projectile-transmitting-energy}
    \end{subfigure}
    \caption{Comparison of (\subref{fig:projectile-transmitting-stress}) average stress and (\subref{fig:projectile-transmitting-energy}) energy transfer as functions of time at different interfaces for MM (solid lines) and monolithic (transparent lines) slabs with transmitting boundary condition on the back-face under projectile impact.}\label{fig:projectile-transmitting-MM-monolithic}
\end{figure}

Boundary conditions bring additional complexities to the performance of these slabs. First, the wave motion is directly affected by the frequency content of the loading, which is a function of the loading type and time scale.
For example, the response of the MM to a projectile impact would be different from that of a blast loading. Second, the response of identical-length MM and comparison slabs will differ due to their different effective wave properties. The effect of the front face loading, \ie PT, PV, PI, and blast, is discussed in \ref{sec:results_loading}. On the other hand, the back-face boundary condition affects the response through reflection and/or transmission at this boundary. It can be traction-free or rigid (fully reflecting), transmitting (non-reflecting), or partially reflective, depending on the impedance of the backing medium. In this section, we compare traction-free and transmitting back-face boundary conditions as two extreme cases. 

Initially, the back-face boundary is set to traction-free. The front face is loaded with the projectile impact. The simulation is run for \qty{200}{\us}. A monolithic slab with the same physical length (10 cells) is simulated with the same conditions, and results are compared with the MM. The average stress and energy transfers interface locations are given in \Figs \ref{fig:projectile-traction-stress} and \ref{fig:projectile-traction-energy}, respectively. Quantities are plotted at the first, second, sixth, and tenth interfaces, corresponding to the impact interface, the first MM interface after the monolithic protective cell, the MM interface in the middle, and the final MM interface before the back monolithic cell.

The average stress results, illustrated in \Fig \ref{fig:projectile-traction-stress}, demonstrate significant differences between MM and monolithic slabs. The monolithic slab has a much higher average stress magnitude throughout the slab, whereas MM has a broader frequency content and lower magnitude stress flux. An investigation into stress peaks unveils that the MM slab has five stress peaks at its initial interface and two within its internal domain. In contrast, the monolithic slab exhibits nine stress peaks at all its interfaces. This pattern of stress peaks is observed to be consistent under various loading conditions, such as prescribed traction with a square stress pulse (not shown here). Furthermore, the analysis of stress dynamics at the impact interface for the MM configuration, denoted by the black curve in Figure \ref{fig:projectile-traction-stress}, showcases a complex contact relationship characterized by multiple peaks and drops during the initial phase of the simulation.

Next, the examination of energy transfer from Figure \ref{fig:projectile-traction-energy} reveals several noteworthy observations. The peak energy transfer level at the terminal interface of the MM slab is approximately \qty{15}{\percent} lower than that of the monolithic slab. Furthermore, the time taken to reach half of the peak energy transfer value at the terminal interface for the MM configuration is approximately five times longer compared to the monolithic structure. This prolonged transmission time signifies the ability of the MM configuration to achieve a significant reduction in wave propagation speed. Additionally, a comparison between the upstream and downstream interfaces highlights the MM's capacity to diminish high-frequency content in the incoming wave. It must be noted that due to the limiting nature of the back face boundary conditions, these observations will be further evaluated for a more practical case. 

Subsequently, referring to Figure \ref{fig:projectile-transmitting-MM-monolithic}, we explore the impact of transmitting back-face boundary conditions. 
The average stress results are shown in Figure \ref{fig:projectile-transmitting-stress}. Comparison with Figure \ref{fig:projectile-traction-stress}
reveals notable 
distinctions between the solutions corresponding to traction-free and transmitting back-face boundary conditions. In the monolithic slab, no tensile stress is observed, given the lack of wave reflection. Conversely, in the MM configuration, tensile stress is observed within the domain, stemming from its inherent microstructural features. Specifically, the presence of free surfaces within the domain leads to the generation of tensile-release waves. 
It is crucial to note that this boundary condition represents an idealized scenario, as reflections would be expected in real-world applications. However, similarly to what was observed earlier, the number of stress peaks is reduced to two from five under the transmitting boundary condition. Another notable observation is that the contact history between the projectile and the MM is noticeably prolonged in the case of the MM configuration when compared to the traction-free scenario. 

The differences between the two slabs are further examined by referring to energy transfers in Figure \ref{fig:projectile-transmitting-energy}. Specifically, the magnitude of energy transfer within the monolithic domain experiences a rapid increase as the elastic wave propagates. In contrast, the MM exhibits a more gradual increase further into the domain. Decrements in the energy transfer correspond to energy flow through the impacted side. The extent of energy attenuation by the slab can be quantified by comparing the energy transfers at the first and final interfaces. The MM attenuates approximately \qty{30}{\percent} of the incoming energy, while the monolithic completely transfers it. Furthermore, akin to our prior findings, the time required to reach the plateau is significantly extended for the MM configuration. These outcomes underscore the potential practical utility of the MM slab.

Given the alignment of main conclusions from the two back-face boundary conditions, \ie slow down of the wave and decay of energy transfer for the MM design, we will use transmitting back-face boundary condition for the remainder of this paper. This approach eliminates the influence of reflections and facilitates the characterization of energy attenuation as the energy curves can reach a plateau.

\begin{figure}[!b]
    \centering
    \includegraphics[width=0.45\textwidth]{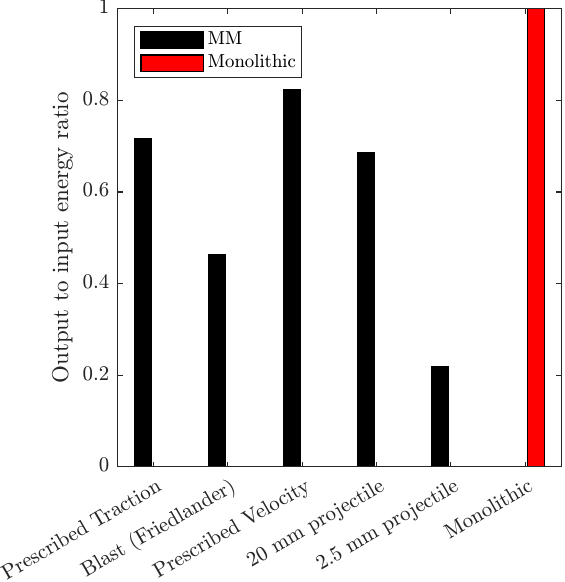}
    \caption{Energy transfer magnitude ratios at the penultimate and incident interfaces.}\label{fig:loading_comparison_energy_ratio_transmitting}
\end{figure}

\subsection{Front face boundary conditions}\label{sec:results_loading} 

The frequency-dependent nature of the MM causes different energy attenuation efficacies under different loading scenarios. To examine the attenuation performance, the back face is set to a transmitting boundary condition, and four different types of loading given in Section \ref{sec:methodology} are applied with the same time scale and stress magnitude at the front face. The ratio of the energy transfer of the last and first interfaces at \qty{200}{\us} for different types of loading is illustrated in \Fig \ref{fig:loading_comparison_energy_ratio_transmitting}. In the figure, the ratio for the monolithic slab for all loading types is represented with a single bar equal to one since there are no wave attenuation mechanisms. For the MM slab, blast and projectile impact loading types significantly have lower energy transfer ratios, around \qtylist[list-units=repeat]{48;68}{\percent}, while ratios for PT and PV ratios are higher at \qtylist[list-units=repeat]{71;82}{\percent}, respectively. The difference in the energy transfer can be attributed to the frequency content of the loading. Moreover, the energy transfer ratio for the projectile impact increases with the projectile length. For instance, the energy transfer ratio for a \qty{2.5}{\mm} projectile impact results in \qty{22}{\percent} transferred energy. The results demonstrate that, unsurprisingly, the energy attenuation is more favorable when the incident frequency content is concentrated around the bandgap frequency.

In conclusion, the MM outperforms the monolithic slab under all loading conditions except for PT, in which it experiences a higher absolute energy transfer at the first interface. However, the energy transfer decay is substantial for all loading conditions, and generally, a lower absolute energy transfer is observed at the downstream interfaces. Moreover,  the wave is slowed down substantially in all cases. In the next section, the comparison is further expanded, considering other types of non-resonant unit cells.

\begin{figure}[t]
    \centering
    \begin{subfigure}{0.34\textwidth}
        \includegraphics[width=\textwidth]{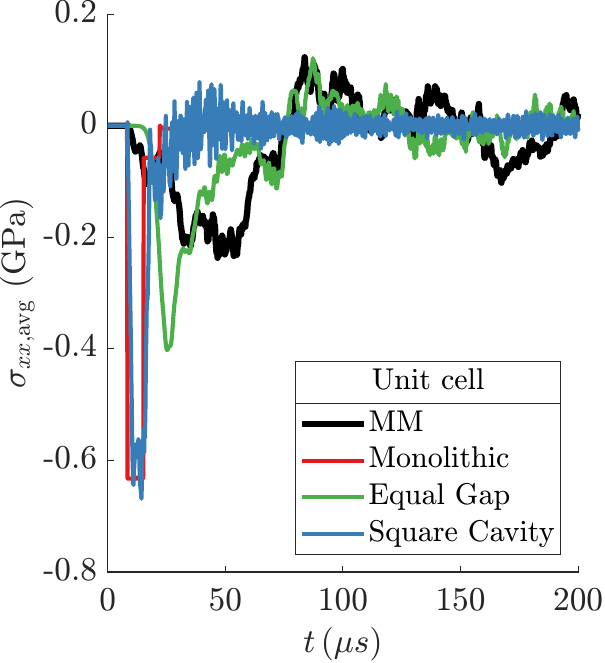}
        \caption{}\label{fig:transmitting-MM-stress}
    \end{subfigure}
    \begin{subfigure}{0.34\textwidth}  
        \includegraphics[width=\textwidth]{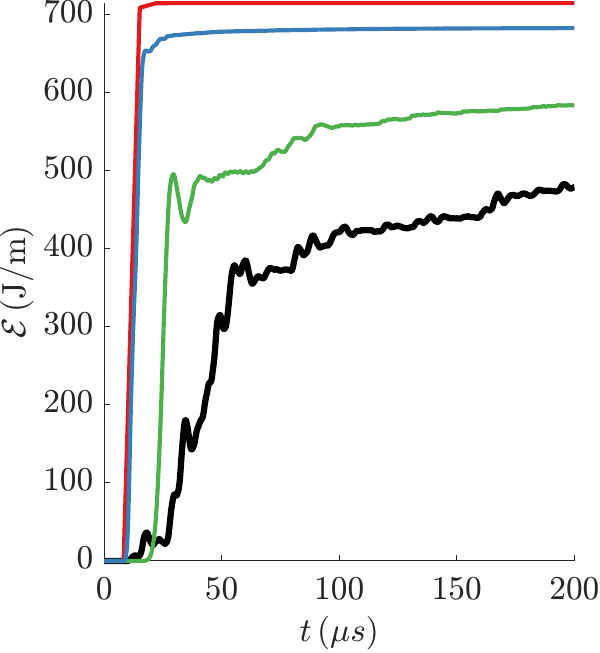}
        \caption{}\label{fig:transmitting-MM-energy}
    \end{subfigure}
    \caption{Comparison of (\subref{fig:transmitting-MM-stress}) average stress and (\subref{fig:transmitting-MM-energy}) energy transfer as functions of time at the final interface for MM and comparison slabs with transmitting boundary condition on the back-face under projectile impact.}\label{fig:transmitting-MM-monolithic}
\end{figure}

\subsection{Comparison slabs}\label{sec:results_comparison} 

This section compares the performance of H-cell arrays with the monolithic, equal gap, and square gap cells given in \Fig \ref{fig:comparison-cells}. Comparisons are conducted with the \qty{30}{\m\per\s}, \qty{20}{\mm} projectile, and transmitting back-face boundary, as explained in the previous sections. All slabs have \num{8} unit cells (or the same thickness in the case of the monolithic slab), excluding protective monolithic cells on both ends. 

The average stress and energy transfer results obtained from the comparison of these unit cells are plotted in \Figs \ref{fig:transmitting-MM-stress} and \ref{fig:transmitting-MM-energy}, respectively. The average stress plot depicts the substantial decrease in the maximum stress magnitude of the MM when compared with the other slabs. The other microstructured slabs also create tensile release waves, similar to the MM. Even though the other comparison cells indicate wave slowdown, we can see that the MM is performing the best in that metric. Furthermore, the energy transfer is significantly lower than the other slabs. The final energy transfer value is around \qty{33}{\percent} lower for the MM than for the monolithic design. In comparison, the energy reduction for the next best design (equal gap) is only around \qty{18}{\percent}. These observations underline the importance of the resonant microstructure compared to simple cellular ones for the mitigation of stress waves. 

\subsection{Heuristic designs}\label{sec:results_heuristic} 

\begin{figure*}[btp]
    \centering
        \begin{subfigure}{0.9\textwidth}
        \includegraphics[width=\textwidth]{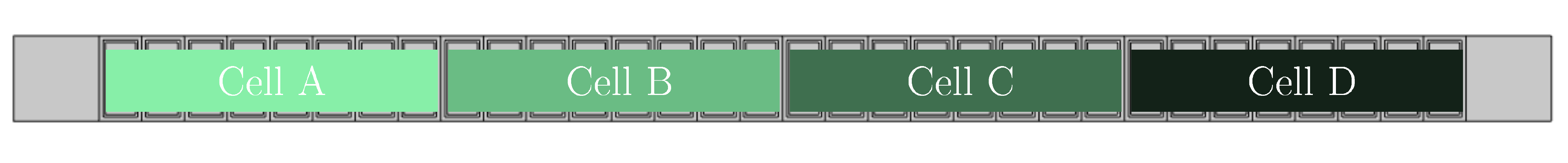}
        \caption{}\label{fig:heuristic_A_D_slab}
    \end{subfigure}
    \begin{subfigure}{0.9\textwidth}  
        \includegraphics[width=\textwidth]{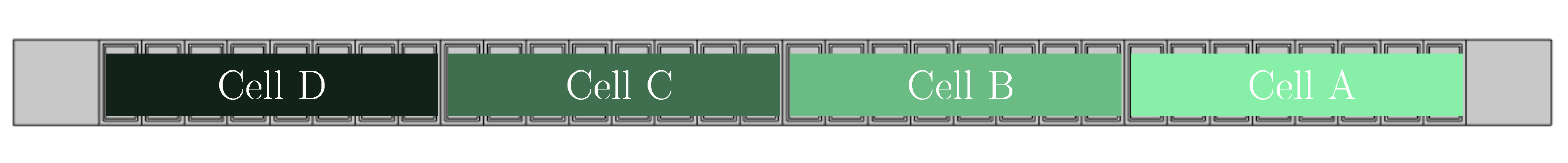}
        \caption{}\label{fig:heuristic_D_A_slab}
    \end{subfigure}
    \caption{Graded slab designs. (\subref{fig:heuristic_A_D_slab}) A to D slab and (\subref{fig:heuristic_D_A_slab}) D to A slab.}\label{fig:heuristic_4x4_slab}
\end{figure*}

\begin{figure*}[t!]
    \centering
    \begin{subfigure}{0.35\textwidth}
        \includegraphics[width=\textwidth]{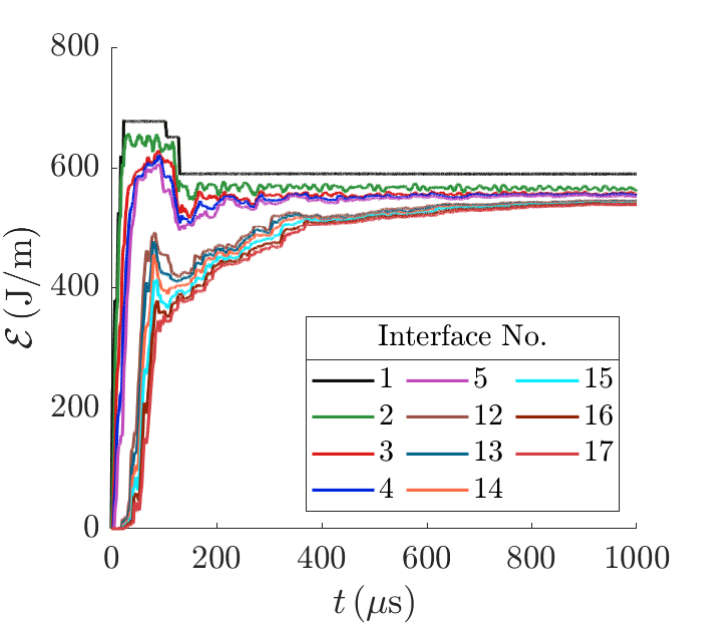}
        \caption{}\label{fig:heuristic_uniform_results}
    \end{subfigure}
    \begin{subfigure}{0.35\textwidth}  
        \includegraphics[width=\textwidth]{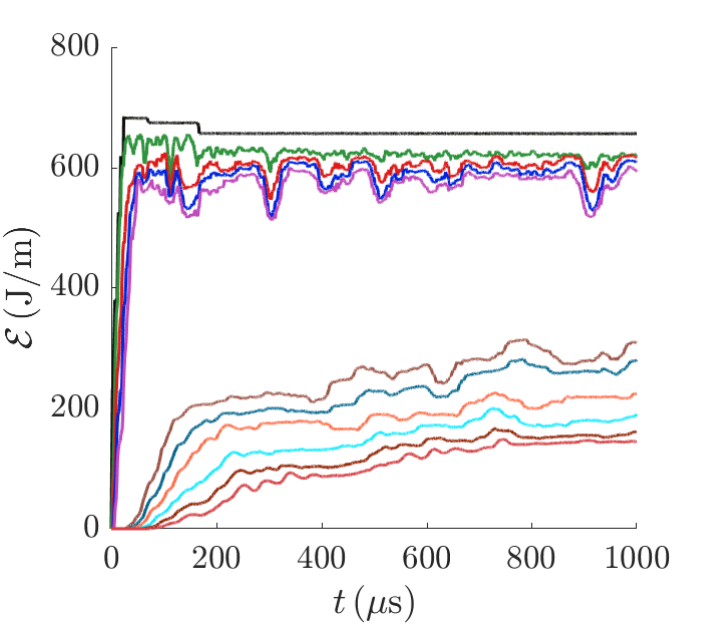}
        \caption{}\label{fig:heuristic_4x4_results}
    \end{subfigure}
    \begin{subfigure}{0.35\textwidth}  
        \includegraphics[width=\textwidth]{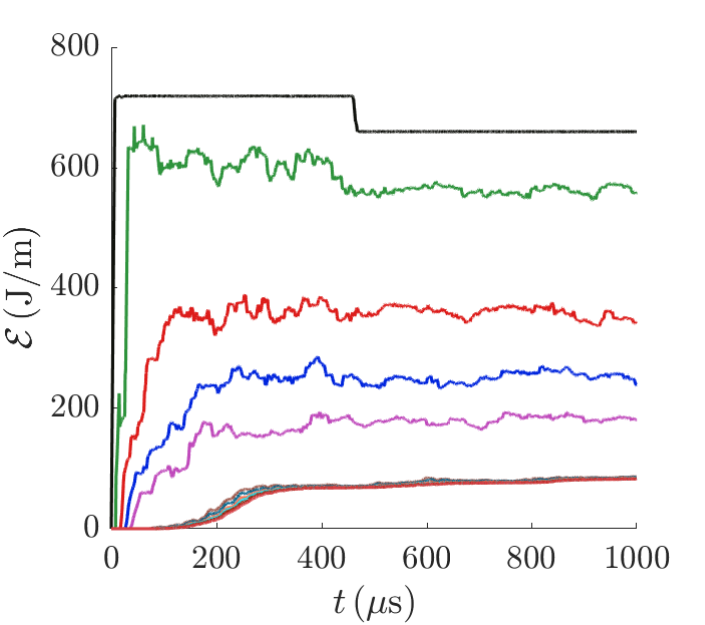}
        \caption{}\label{fig:heuristic_4x4_flipped_results}
    \end{subfigure}
    \begin{subfigure}{0.33\textwidth}  
        \includegraphics[width=\textwidth]{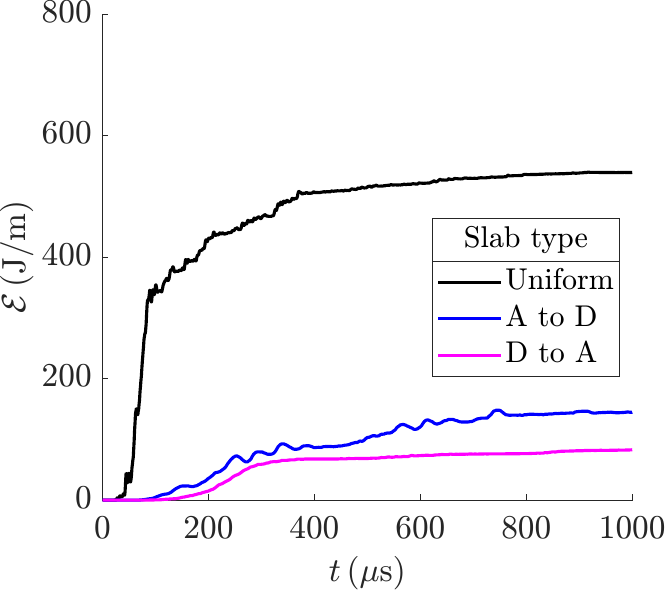}
        \caption{}\label{fig:heuristic_final_interface}
    \end{subfigure}
    \caption{Energy transfer comparison for the uniform and graded designs: (\subref{fig:heuristic_uniform_results}) uniform slab, (\subref{fig:heuristic_4x4_results}) A to D, (\subref{fig:heuristic_4x4_flipped_results}) D to A. Only the first five and the last six interfaces are plotted. (\subref{fig:heuristic_final_interface}) Energy transfer magnitude at the final interface for different slabs.}\label{fig:heuristic_results}
\end{figure*}

In this section, possible nonuniform (graded) MM arrays are explored to further improve the impact response of the structure. A \num{16} cell MM slab has been modified by changing the material properties after every four cells, thus resulting in a four-by-four structure shown in \Fig \ref{fig:heuristic_A_D_slab}. In practice, it is far more feasible to create such gradation through geometric modifications. However, for the sake of computational consistency and speed, the concept is tried here using density and modulus modifications. To study directionality considerations, the same slab has been flipped and analyzed, as shown in \Fig \ref{fig:heuristic_D_A_slab}. To showcase the performance difference, the initial four-cell group (A) has kept the same properties as alumina, and the following groups' densities and Young's moduli are modified to have bandgaps centered around two times lower bandgap frequencies progressively:

\begin{align}
    \rho_{\mathrm{A}} = \frac{\rho_{\mathrm{B}}}{2} = \frac{\rho_{\mathrm{C}}}{4} = \frac{\rho_{\mathrm{D}}}{8}, \\
    E_{\mathrm{A}} = 2E_{\mathrm{B}} = 4 E_{\mathrm{C}} = 8E_{\mathrm{D}}.
\end{align}

Consequently, the wave speed is also modified and decreased by a factor of 2 each time from A to B to C to D. On the other hand, the impedances are equal to that of the original material, that is for cells A.

The energy transfer plots for uniform material, A to D, and D to A graded slabs are given in \Figs \ref{fig:heuristic_uniform_results}, \ref{fig:heuristic_4x4_results}, and \ref{fig:heuristic_4x4_flipped_results}, respectively. The energy ``capture'' capacity of the uniform array quickly stagnates after only a few cells, and the downstream cells do not engage efficiently with the wave energy arriving at those locations. However, the A to D array is more robust in interacting with the wave, as apparent by the gaps between energy transfer at the interfaces. This is caused by the lower frequency bandgaps of the downstream groups. On the other hand, when this design is flipped, \ie the D to A array, most of the energy transfer attenuation is achieved in the first group. Downstream groups, C to A, do not have a significant engagement with the incoming wave.

The final interface energy transfer comparison of the three designs is given in \Fig \ref{fig:heuristic_final_interface}. The effect of the difference in wave speed of the materials can be clearly seen in the wave arrival time. When the heavier and slower cell is in the front, the energy transfer magnitude is lower, and the wave arrival time is longer. The terminal energy magnitude of the D to A slab is \num{6.5} times lower than the uniform slab. These are particularly encouraging results that show the impact of graded MM design. Further investigation is needed to better understand the possible geometrical designs to target this performance.

\subsection{Damping}\label{sec:results_damping} 

Material damping is a critical mechanism for energy attenuation. Especially for MMs, damping can be used as an advantageous mechanism to reduce energy transfer further. For this reason, material damping is added to the linear finite element model of the MM and monolithic slabs, and the behavior is explored for different damping constants. 

Material damping in finite element modeling can be treated with mass-proportional or stiffness-proportional damping. LS-DYNA calculates internal forces by integrating stresses over the element area. Damping forces are added to these forces by \cite{kumar_damping_2021},
\begin{equation}
\mathbf{a}^n=\mathbf{M}^{-1}\left(\mathbf{P}^n-\mathbf{F}^n-\mathbf{F}_{\text {damp }}^n\right),
\label{eq:EOM_wDamping}
\end{equation}
where $\mathbf{a}^n$ is the acceleration vector, $\mathbf{M}$ is the mass matrix, $\mathbf{P}^n$ and $\mathbf{F}^n$ are the external and internal load vectors, respectively, and  $\mathbf{F}_{\text {damp }}^n$ is the damping force, calculated by \cite{lstc_keyword_2017},
\begin{equation}
\mathbf{F}_{\text {damp }} = \sum D_s m {\vc v} = \sum 4\pi f \zeta m {\vc v},
\label{eq:dampingForce}
\end{equation}
where $D_s$ is the damping constant, $m$ is the node's mass, $\vc{v}$ is the node's velocity vector, $\zeta$ is the critical damping ratio or material damping factor, and $f$ is the damping frequency. The summation $\sum$ refers to the assembly of nodal forces to 
$\mathbf{F}_{\text {damp }}$.
 Another constant that describes damping is the quality factor or internal friction, $Q^{-1}=2\zeta$ \cite{liu_damping_2020}. Deduced from the reported $Q^{-1}$ in \cite{lambrinou_elastic_2007}, the critical damping ratio for alumina at room temperature is  $\zeta 
= $ \num{2.5e-5}. 

Next, we relate the damping force to the Rayleigh damping.
Damping can be modeled using classical Rayleigh damping, where the damping matrix, $\mathbf{C}$, is calculated by a linear combination of mass and stiffness matrices, $\mathbf{M}$ and $\mathbf{K}$, respectively \cite{kumar_damping_2021},
\begin{equation}
    \mathbf{C} = \alpha \mathbf{M} + \beta \mathbf{K},
\end{equation}
where $\alpha$ and $\beta$ are the Rayleigh damping coefficients. The contribution from the $\alpha$ term is equivalent to 
the damping force in 
$D_s$. More specifically, $\alpha =  2\pi f \zeta$, where $f$ is the frequency for which the value of $\zeta$ is measured. This corresponds to $f \approx $ \qty{10}{\kHz} in \cite{lambrinou_elastic_2007}, resulting in a baseline $\alpha$ of around $1.6 / \mathrm{s}$. 
As for an upper bound for $\alpha$, the critical damping ratio for alumina with a void ratio of 2\% can go up to $\zeta$ = \num{0.01} \cite{panteliou_damping_2009}. This is a realistic upper bound to consider for 3D printed ceramic as depending on the printing parameters, such void ratios can be observed. The corresponding $\alpha$ to this upper bound $\zeta$ is around $600 / \mathrm{s}$. 
As there is an ambiguity in the actual damping parameter of the ceramic, subsequent analysis uses $\alpha$ values ranging from around \num{e-2} to \num{e6}$/\mathrm{s}$. This not only contains the range of values reported based on the results in \cite{lambrinou_elastic_2007} and \cite{panteliou_damping_2009} but also can examine how the response is affected by lower and higher damping values.

\begin{figure}[t]
    \centering
        \begin{subfigure}{0.35\textwidth}
        \includegraphics[width=\textwidth]{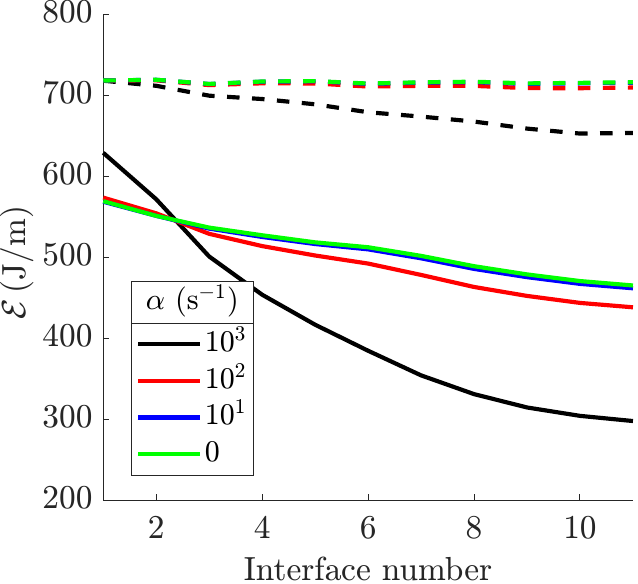}
        \caption{}\label{fig:damping-interfaces}
    \end{subfigure}
    \begin{subfigure}{0.35\textwidth}  
        \includegraphics[width=\textwidth]{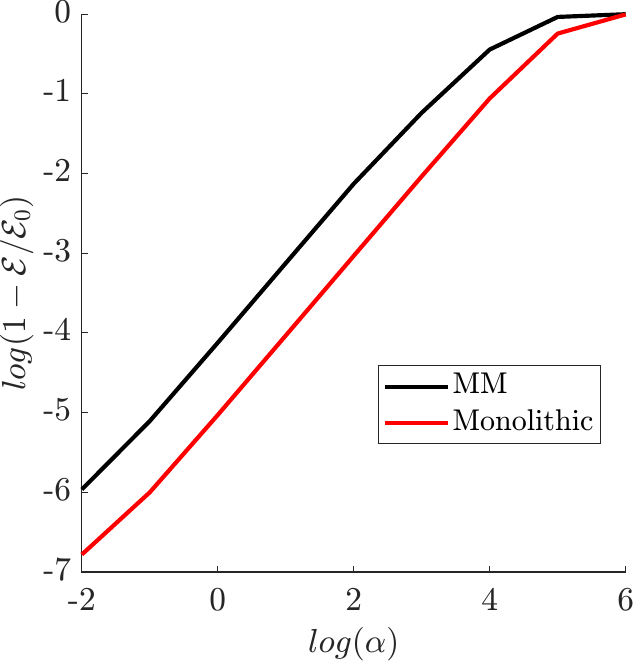}
        \caption{}\label{fig:damping-log-log}
    \end{subfigure}
    \caption{(\subref{fig:damping-interfaces}) Energy transfer comparison for different damping constant values. Solid lines represent the MM, while dashed lines represent the monolithic slabs. (\subref{fig:damping-log-log}) Normalized difference of damped and lossless energy transfers plotted against the damping constant ($\mathrm{s^{-1}}$) for the MM and monolithic slabs.}\label{fig:damping-results}
\end{figure}

Energy transfer is compared at different interfaces for varying $\alpha$ and for MM and monolithic designs in \Fig \ref{fig:damping-interfaces}. From the same projectile impact, the MM and the monolithic slabs differ at the first interface for the lossless case due to the effective impedance of the slabs. What stands out in this figure is the vast disparity between the energy transfer difference at the final interface for the same damping constant, \ie $\alpha = $ \num{e2}$/\mathrm{s}$. The energy transfer difference with the lossless system becomes significant for damping constants higher than \num{e2}$/\mathrm{s}$. Even though the reported damping constant of alumina from \cite{lambrinou_elastic_2007}
is smaller than this value, as discussed above, larger $\alpha$ can be encountered for 
higher void ceramics \cite{panteliou_damping_2009}.
Likewise, changing to a different material with a higher damping constant may also take advantage of lower stress magnitude and energy transfer.

To compare the effect of different damping coefficients for the MM and monolithic slabs, the energy transfer at the final interface for the damped slabs ($\mathcal{E}$) is subtracted from the lossless system ($\mathcal{E}_0$)
and normalized with respect to $\mathcal{E}_0$. 
A value close to zero ($\mathrm{log}(1 -\mathcal{E}/ \mathcal{E}_0) \rightarrow -\infty$) means that the lossy system provides no advantage in further energy decay relative to the lossless one, while as this ratio tends to one 
($\mathrm{log}(1 -\mathcal{E}/ \mathcal{E}_0) \rightarrow 0$)
the damping has a significant effect and reduces the relative energy transfer to zero. The normalized energy ratio of both slab designs is plotted against $\alpha$ in \Fig \ref{fig:damping-log-log}. The slopes of the log-log curves are equal to one. A one-dimensional analysis, not presented here for brevity, confirms this slope for the monolithic design.
The energy transfers in both cases are getting closer to zero for high damping constants. For the lower damping constants, there is a constant difference between the MM and monolithic that can be attributed to the higher structural damping capacity of the MM. That is, there is even an added benefit relative to a monolithic material response in terms of the energy decay of the MM design when material damping is considered.
This higher potential can be exploited for energy mitigation applications such as impact and blast protection.

\section{Summary and conclusions}\label{sec:summary}

In this work, a locally resonant elastic microstructured material is analyzed for impact mitigation using the finite element method. The frequency-dependent response is captured by time domain simulations, incorporating boundary effects for planar slabs with finite thickness. 

First, it was shown that the number of cells in the MM slab is of little importance after only (in this case) eight cells. Changes in the frequency content of the propagating wave inhibit further energy flux reduction. The performance of the slab can be improved by varying the cell design to broaden the performance frequency range. We showed that the graded design can remarkably enhance the energy flux attenuation of the slab.

Second, the nature of the loading affects the response drastically. For example, the MM can proportionally attenuate more energy from shorter projectile impacts. Nevertheless, for all the loading cases considered, the MM slab exhibited a higher reduction in normalized energy transfer compared to the monolithic and comparison slabs. 

Finally, it was shown that the performance of the MM slab also exceeds the monolithic and other non-resonant comparison slabs in terms of peak energy transfer and wave slowdown. The peak energy transfer at the final interface of the MM slab can be up to 78\% and 60\%  
lower than those of the monolithic and equal gap (best of comparison) slabs.
Similarly, the MM slab can delay the arrival time of the wave five times more while reducing the number of stress peaks inside the domain. Besides, damping in the structure can significantly increase the energy transfer mitigation, up to four times better than a monolithic slab. These findings highlight the potential usefulness of MMs in impact and blast protection. 

Further research should be undertaken to investigate the nonlinear response of the structure. Material damping and failure account for energy dissipation while drastically changing the stress wave profile. Our preliminary results, accounting for material failure (to be presented in an upcoming publication), indicate that the MM slab significantly improves the energy transfer attenuation compared to the monolithic slab, even in the presence of widespread failure. 
Further analysis focusing on 2D wave propagation in MM structures has also underlined the potential to leverage anisotropic wave speed with designs developed using generative neural networks to further improve impact mitigation using microstructured media \cite{wang_generative_2024}.

\section{Acknowledgements}
The authors wish to thank DEVCOM Army Research Laboratory for continued support throughout this effort. This research was supported by DEVCOM Army Research Laboratory through Cooperative Agreement W911NF--20--2--0147, with additional support from The University of Tennessee. The authors would also like to thank the University of Tennessee Infrastructure for Scientific Applications and Advanced Computing (ISAAC). A portion of the computation for this work was performed with ISAAC computational resources.

\bibliographystyle{unsrtm}
\biboptions{sort&compress}

\end{document}